\documentclass[12pt,preprint]{aastex}

\usepackage{epsfig}

\usepackage[outdir=./]{epstopdf} 

\shorttitle{Fulleranes and the 21\,$\mu$m feature}
\shortauthors{Zhang}

\begin{document}

\title{Are fulleranes responsible for the 21 micron feature?
}

\author{Yong Zhang}
 \affil{School of Physics and Astronomy, Sun Yat-sen University,  2 Daxue Road,  Tangjia, Zhuhai,  
Guangdong Province,  China}
 \affil{Laboratory for Space Research, Faculty of Science, The University of Hong Kong, Pokfulam Road, Hong Kong, China}
\email{zhangyong5@mail.sysu.edu.cn}

\begin{abstract}

Recent detections of C$_{60}$, C$_{70}$, and C$_{60}^+$ in space induced extensive studies of 
fullerene derivatives in circumstellar environments. As the promising fullerene sources,
protoplanetary nebulae (PPNe) show a number of unidentified bands in their infrared spectra,
among which a small sample exhibits an enigmatic feature at $\sim21$\,$\mu$m. Hydrogenation
of fullerenes can produce fulleranes emitting new infrared bands. In this paper, we investigate
the possibility of fulleranes (C$_{60}$H$_m$) as the carrier of  the 21\,$\mu$m feature in terms of
theoretical vibrational spectra of fulleranes. The evidences favoring and disfavoring
the fullerane hypothesis are presented. We made an initial guess for the hydrogen coverage
of C$_{60}$H$_m$ that may contribute to the 21\,$\mu$m  feature.

\end{abstract}


\section{Introduction}

The `21\,$\mu$m' feature refers to an infrared emission band peaking at 20.1\,$\mu$m
that was first discovered by \citet{kv89} in four circumstellar envelopes of
evolved stars.
This feature is rare compared to other circumstellar dust features. Thus far, it has been detected 
toward only 27  evolved stars, including 18 in the Galaxy and 9 in the Large and Small
Magallenic Clouds  \citep{ml16}.  
The 21\,$\mu$m sources have some properties in common. Their optical-infrared spectra reveal
two blackbody components, corresponding to radiation from the stellar photosphere and the dust shell.
This is an evidence that stellar outflows have terminated and the envelope has been detached from
the central star. Therefore, they represent a short evolutionary stage between the asymptotic giant branch 
(AGB) and the planetary nebula (PN), usually denoted as protoplanetary nebula (PPN). All the 21\,$\mu$m sources
are carbon rich, exhibiting absorption features from C$_2$, C$_3$, and CN \citep{ba95} 
as well as the aromatic C-H bands at 3.3 and 11.3\,$\mu$m \citep{hr08}.
Moreover, a broad emission band around 30\,$\mu$m always appears along with
the 21\,$\mu$m feature \citep{hr08}. The identificaton of the 30\,$\mu$m feature
is another unsolved problem \citep[see][for a review]{jz10}.
It is worth noting that the spectra of
two supernova remnants show a strong dust feature exactly peaking at 21\,$\mu$m \citep{rho18}.
Although its peak wavelength slightly differs from the 21\,$\mu$m feature, it would be interesting
to investigate whether their carriers belong to a similar molecular family.

Identification of the 21\,$\mu$m feature is vital for understanding circumstellar chemistry and
matter cycle in galaxies.  Although  a  number of candidate materials have been proposed as
its carrier \citep[see][and the references therein]{jz10}, no consensus emerges. Some of
them can be rejected. \citet{zj09} found that S-, Si-, and Ti-containing compounds are unlikely 
responsible for the 21\,$\mu$m feature as a cause of low abundances of these elements. The strength
of the 21\,$\mu$m  feature suggests that its carrier is composed of rich elements.
Fe oxides can be ruled out because they would emit too broad 21\,$\mu$m feature or
some subfeatures that were never detected \citep{zj09,li13}.  A carrier candidate
that has not been adequately examined by these authors is hydrogenated fullerenes (fulleranes).

Since the discovery of C$_{60}$ \citep{kroto},  fullerenes or their derivatives have been
long conjectured to be ubiquitous throughout interstellar and circumstellar space. This was
subsequently confirmed by the detection of C$_{60}$ and C$_{70}$ in PNe
\citep{cami10,gar10} and the convincing assignment of a few diffuse interstellar bands as 
C$_{60}^+$ \citep{cam}. 
C$_{60}$ can be rapidly formed in PPN stage \citep{zhang11}. \citet{gar10} suggested that the
environments of forming fullerenes could be hydrogen-rich.  Fullerenes have high proton affinities.
When mixed with atomic hydrogen, C$_{60}$ can be efficiently hydrogenated into fulleranes
in laboratory \citep{ci09,ig12}.
Given the high stability of fullerene structure as well as the high abundance of cosmic carbon and hydrogen,  
it is attractive to investigate the vibrational spectra of fulleranes and search for their existence 
in circumstellar envelopes.

Based on a simple force-field model, \citet{web95} found that fulleranes can radiate a broad band
in the wavelength range from 19--23\,$\mu$m, and raised the possibility of fulleranes as 
the carrier of the 21\,$\mu$m feature. 
In laboratory environments, only specific fulleranes can be synthesized, making it hard to verify
this hypothesis.  Recently, more accurate calculations of vibrational spectra can be performed
with the development of computational chemistry methods and computing facilities,
allowing us to re-examine whether fulleranes can account for the 21\,$\mu$m feature.
This motivates the current paper.

\section{Computations}

In order to test the proposal of C$_{60}$H$_{m}$ as the carrier of the 21\,$\mu$m feature,
we calculated the vibrational spectra of selected C$_{60}$H$_{m}$  with  even number of hydrogen atoms
\citep[see][for the details]{zhang17}. 
It is computationally impossible to derive the  spectra of all C$_{60}$H$_{m}$ because of  the
enormous isomer number.  
In laboratory conditions, C$_{60}$ can be readily hydrogenated into C$_{60}$H$_{36}$ and then 
 form C$_{60}$H$_{18}$ through thermal annealing \citep[e.g.][]{ig12}. With increasing hydrogen
coverage, the carbon hybridization goes from $sp^2$ to $sp^3$. This reduces the stability of
the carbon cage. Moreover, fulleranes are exposed to ultraviolet (UV) photons which are abundantly 
present in PPNe, and thus undergo dehydrogenation. As a result, C$_{60}$H$_{m}$ with large $m$ number 
is unlikely exist in PPNe.  A total of 55 isomers belonging to 11 C$_{60}$$m$ species 
($m=2,4....20$, and $36$) were selected for the computations. These isomers
should represent those with lowest energy and thus being the most stable ones for a given $m$ number.
For computational convenience, we did not consider fulleranes with
odd number of hydrogen atoms and their ionized counterparts although they are
possibly present in astronomical objects. A theoretical investigation of those species  will be 
the subject of a follow-up paper.

We performed density functional theory (DFT) calculations using the B3LYP  and BH\&HLYP hybrid functionals 
in combination with polarization consistent basis set PC1. The  vibrational frequencies were
obtained using a double-scaling-facter scheme \citep{lc12}. This method can well reproduce the
experimental spectrum of C$_{60}$ \citep{zhang17}.
Assuming that C$_{60}$H$_{m}$ infrared spectra are thermally excited in circumstellar environments, 
we derived the fluxes through scaling the computed intrinsic strengths by a Boltzmann factor
with a temperature of $300$\,K. A Drude profile with a fixed width of 0.3\,$\mu$m was convolved to broaden 
the features, for the purpose of comparing the computed spectra with the observed ones. Finally,
the spectrum of each C$_{60}$H$_{m}$ was obtained by co-adding its isomer spectra.

\section{Results and discussion}

\begin{figure*}
\epsfig{file=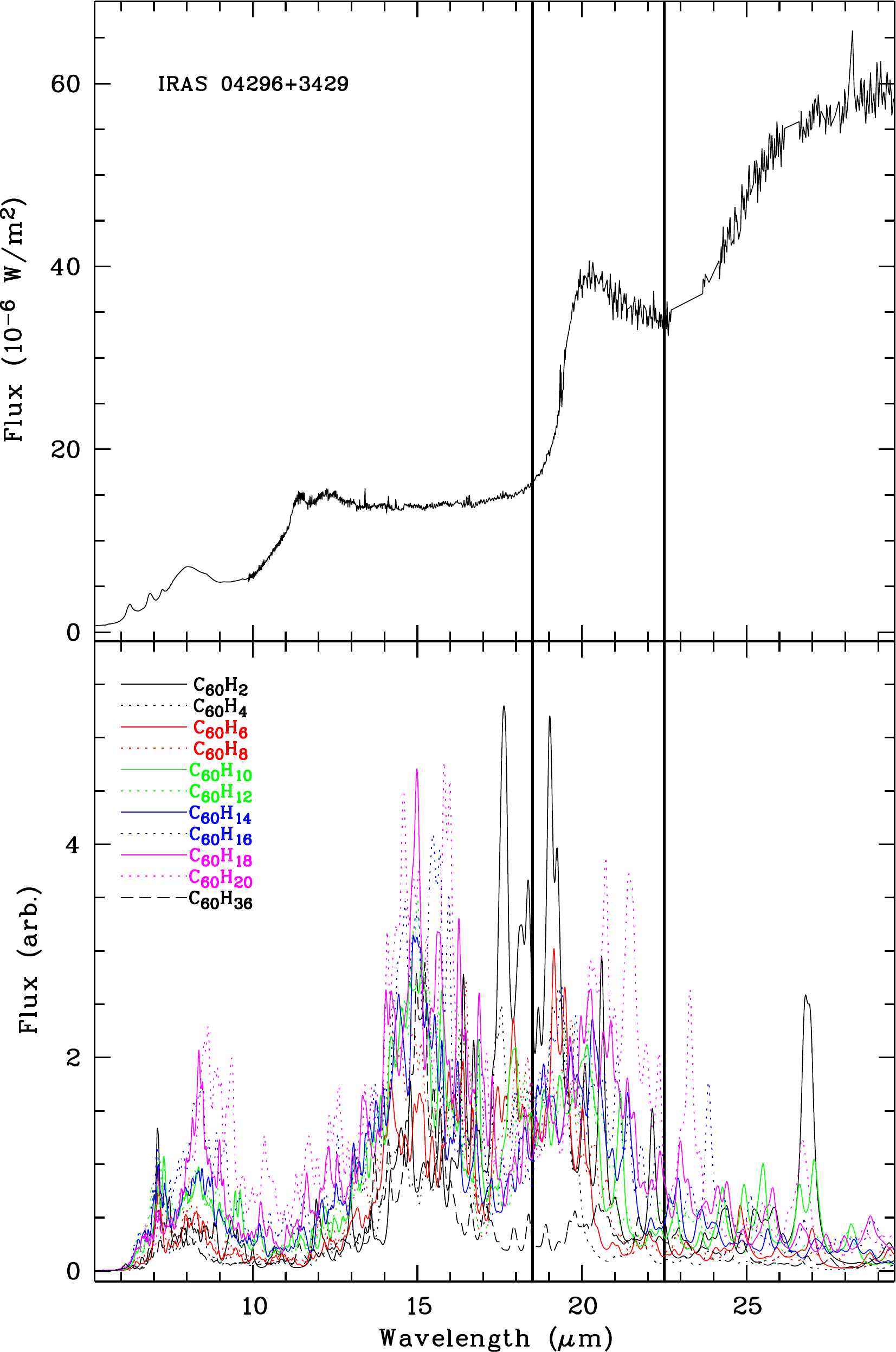, height=20cm}
\caption{The {\it Spitzer}/IRS  spectrum of the PPN IRAS 04296+3429 
(upper panel) and the theoretical spectra of C$_{60}$H$_m$ (lower panel). 
Note that the continuum has not been subtracted for the observational spectrum.
The vertical lines mark the wavelength range of the observed 21\,$\mu$m band.}
\label{spe}
\end{figure*}

The theoretical spectra of C$_{60}$H$_{m}$ in the 5--30\,$\mu$m range are presented in Figure~\ref{spe}. 
Although the vibrational modes are spread out over a wide wavelength range, strong emission features
mainly concentrate between 5--10, 13--17, and 18--23\,$\mu$m. 
The  3--4\,$\mu$m wavelength range, where the C-H stretching modes can produce strong features, has been
extensively discussed in a previous paper \citep{zhang17}, and thus is not included in this figure.
For comparison, this figure also shows the observational spectrum of a PPN IRAS 04296+3429, which 
was taken from the {\it Spitzer} archive \citep{zhang10}. As shown in Figure~\ref{spe},
a strong 21\,$\mu$m emission band exists in the spectrum of IRAS 04296+3429, which is well
confined in the wavelength range of 18--23\,$\mu$m. It is clear that a weighted combination of 
different C$_{60}$H$_{m}$ spectra provides a potential to reproduce the 21\,$\mu$m feature. 
The features grouping in the wavelength of 5--10\,$\mu$m are also detected in the observational
spectrum, providing a futher support for C$_{60}$H$_{m}$ as the carrier of the 21\,$\mu$m feature.
However, no strong feature in the 13--17\,$\mu$m range is detected in the observational spectrum, but
which instead exhibits features in 10--13\,$\mu$m range.
The 10--13\,$\mu$m features are not seen in 
the theoretical spectra of C$_{60}$H$_{m}$, and usually are attributed to silicon carbides
in the literature.

Any single C$_{60}$H$_{m}$ spectrum is unable to reproduce the observed 21\,$\mu$m feature. Fulleranes, if being responsible for the 21\,$\mu$m feature in PPNe, should be a mixture of various isomers with different hydrogenation degrees. It is natural to expect that their spectral pattern 
may vary from source to source. However, astronomical observations have shown that  
the 21\,$\mu$m features have a remarkably consistent profile \citep{vol99}. The same
problem exists for the mixtures of polycyclic aromatic hydrocarbon molecules as the carrier
of unidentified infrared emission bands, and has been extensively discussed 
\citep[see][for the details]{kz13}.  In order to explain the observations, we conjecture that
only certain fullerane family can be formed and survive in the exclusive PPN environment.
If the UV radiation is absent, hydrogen is mainly in the molecular state, and thus
hydrogenation of C$_{60}$  is unlikely to occur; if it is too strong, it may substantially
dehydrogenate fulleranes, or even break the carbon cage.
The rigorous environments required for the existence of fulleranes are compatible with
the rareness of the 21\,$\mu$m feature in astronomical objects.

\begin{figure*}
\epsfig{file=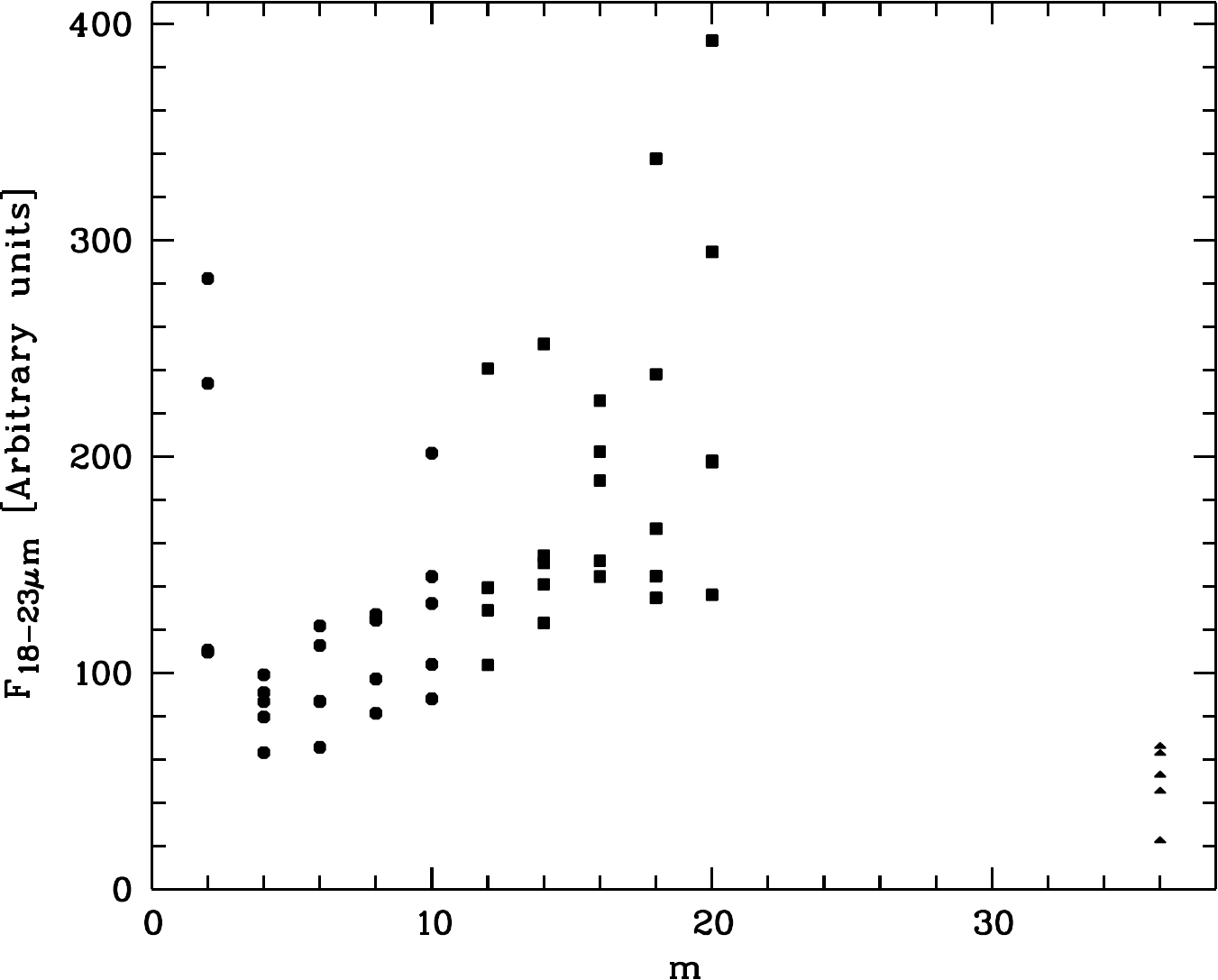, height=10cm}
\caption{Intrinsic strengths of the C$_{60}$H$_{m}$ features lying
in the wavelength range
of 18--23\,$\mu$m versus the $m$ values.
The filled circles, squares,
triangles represent the slightly ($m=$2--10), moderately ($m=$12--20), and heavily
($m=$36) hydrogenated C$_{60}$, respectively.
}
\label{21vsm}
\end{figure*}

Figure~\ref{spe} indicates that C$_{60}$H$_{36}$ 
can be ruled out as the carrier the 21\,$\mu$m feature
due to the lack of features lying between the 18--23\,$\mu$m wavelength range.
C$_{60}$H$_m$ with $m=$2--8 is unlikely responsible for 
the 21\,$\mu$m feature because their strong features have peak wavelengths shorter than
20\,$\mu$m. Moreover,
C$_{60}$H$_{2}$ can produce a strong feature around 27\,$\mu$m, which has never
been detected. 
Figure~\ref{21vsm} presents the contributions from each C$_{60}$H$_m$ isomer
to the total intrinsic strength of the features lying in the
18--23$\mu$m range. Except C$_{60}$H$_{2}$
and C$_{60}$H$_{36}$, the strengths increase with increasing $m$ values,
suggesting that moderately hydrogenated C$_{60}$ among fulleranes
produce the strongest vibrational bands
in the  wavelength range encompassing the 21\,$\mu$m feature.

\begin{figure*}
\epsfig{file=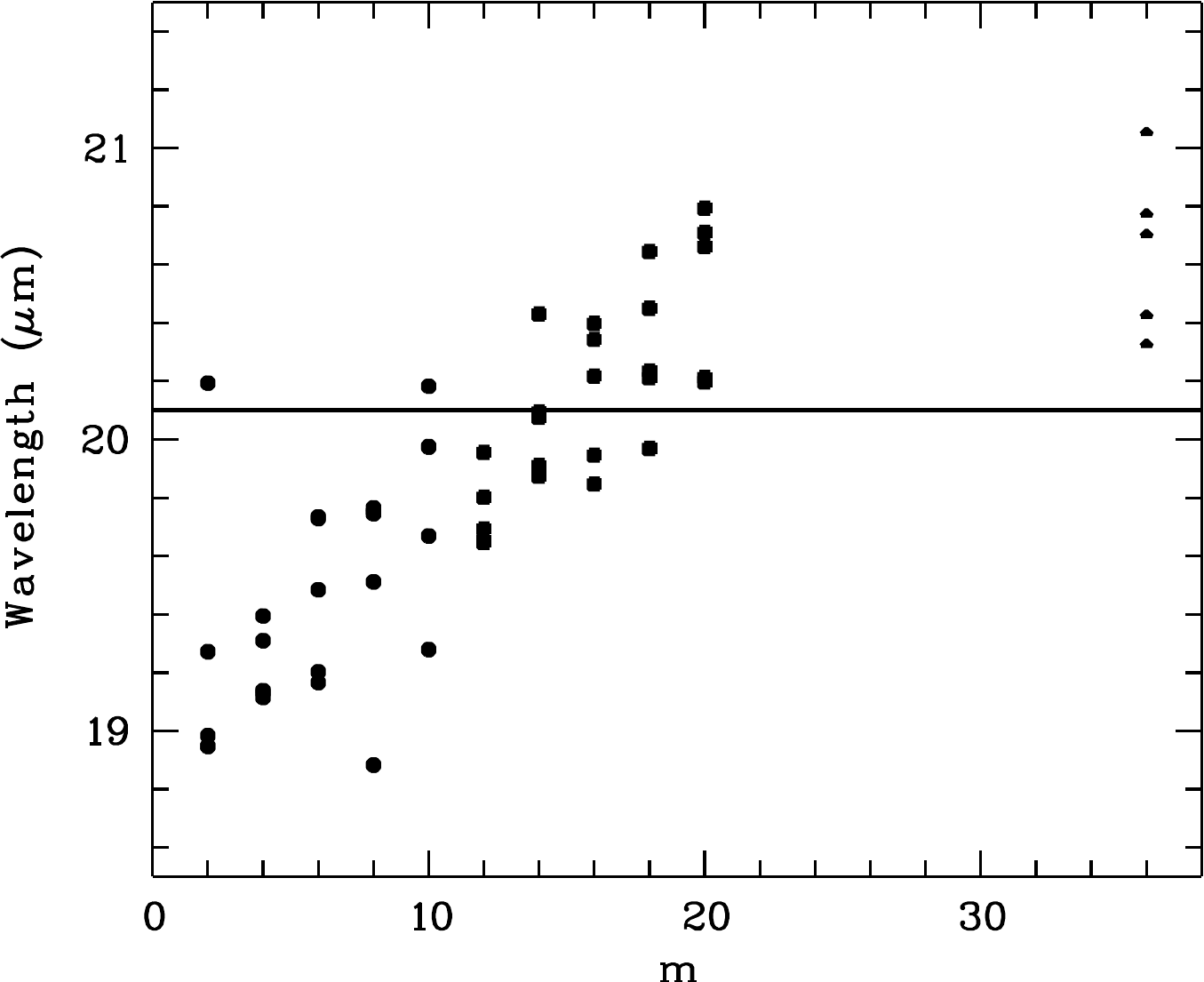, height=10cm}
\caption{The intensity-weighted wavelengths of the features lying in the wavelength range of
18--23\,$\mu$m versus the $m$ values of  C$_{60}$H$_{m}$. 
Symbols are the same as those in Figure~\ref{21vsm}. The horizontal line denotes the wavelength
position of the 21\,$\mu$m feature.}
\label{wave}
\end{figure*}

In Figure~\ref{wave}, we examine the intensity-weighted wavelengths ($\lambda_{21}$)
of C$_{60}$H$_m$ as a function of $m$, which is defined as
\begin{equation}\label{eq1}
\lambda_{21}=\frac{\sum\limits_i \lambda_iF_i}{\sum\limits_i F_i}
\end{equation}
for $18\,\mu m\le\lambda_i\le23\,\mu m$,
where $F_i$ is the intrinsic strength of the $i^{\rm th}$ mode at the wavelength of $\lambda_i$.
An inspection of Figure~\ref{wave} reveals that there is a approximately linear
 trend of longer intensity-weighted wavelengths with increasing hydrogenation degrees.
This is in agreement with the predictions by the force-field model  of
\citet{web95}. Comparing with the observed peak wavelength (20.1\,$\mu$m),
we infer that moderately hydrogenated fulleranes with $10<m<20$
are the promising carrier for the 21\,$\mu$m feature.

\begin{figure*}
\epsfig{file=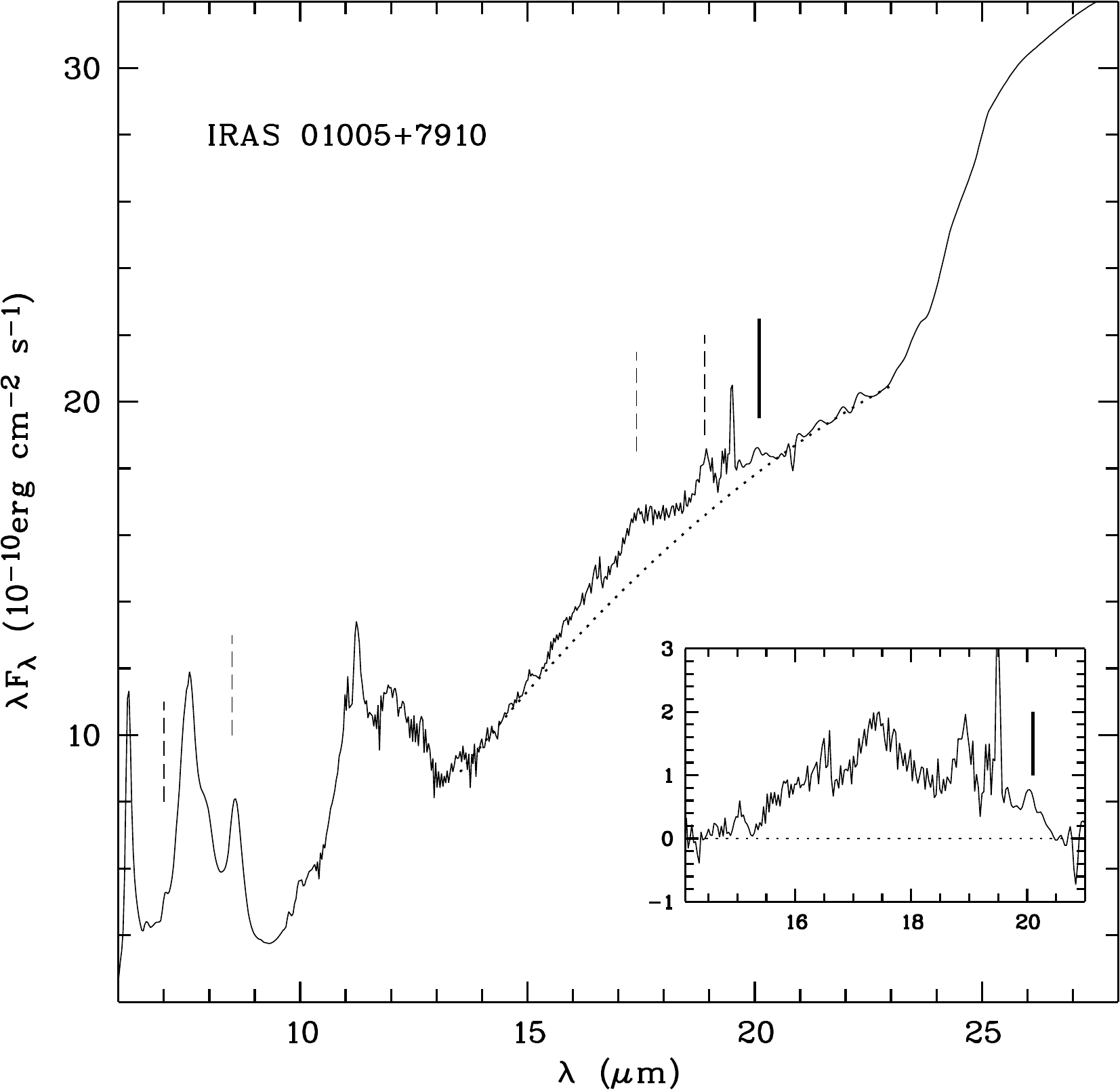, height=12cm}
\caption{The {\it Spitzer}/IRS  spectrum of the PPN IRAS 01005+7910. The vertical solid and
dashed lines mark the positions of the 21\,$\mu$m feature and the four C$_{60}$ bands, respectively.
The lower-right panel shows the continuum-subtracted spectrum.
}
\label{I01005}
\end{figure*}

While the existence of fulleranes in astronomical environments seems plausible, there is as yet no unambiguous 
detection.  \citet{dg16}  failed to detect the C-H stretching bands at 3.4--3.6\,$\mu$m 
in two PNe exhibiting strong C$_{60}$ emission, suggesting that fulleranes 
might have been destroyed by strong UV radiation or mostly ionized.
Based on a comparison between the laboratory spectrum of gaseous C$_{60}$H$^+$ 
and the observations, \citet{pa19} speculated that C$_{60}$H$^+$ might contribute
to the spectra of two C$_{60}$-containing PNe. Presumably,
exposed on very strong UV radiation, moderately and heavily hydrogenated fulleranes 
cannot survive in PNe. This is compatible with the non-detection of the 21\,$\mu$m feature in PNe.
\citet{zhang13} reported a tentative detection of fulleranes 
in the C$_{60}$ source IRAS 01005+7910, which is a PPN about to enter the PN stage.
 IRAS 01005+7910 is not assigned as a 21\,$\mu$m source in the previous literature.
However, a closer view of its infrared spectrum clearly reveals a faint feature exactly
peaking at 20.1\,$\mu$m (Figure~\ref{I01005}). This feature is much weaker than
the C$_{60}$ bands, and has not been noted previously.
If attributing it to fulleranes, we can hypothesize that
fulleranes in this PPN are undergoing dehydrogenation, resulting in a transition
from moderately hydrogenated  C$_{60}$ to slightly and none hydrogenated  C$_{60}$.
The enrichment of slightly hydrogenated  C$_{60}$ may partly explain the 15--20\,$\mu$m plateau 
emission in the spectrum of IRAS 01005+7910.

A criticism of the fullerane hypothesis was advanced by \citet{po04}, who pointed out that
the actual wavelength of the C$_{60}$ feature is not coincident with that of
the emission band in the meteoritic nanodiamonds shown by \citet{hi98}. The argument stems
from the fact that nanodiamonds and fulleranes have a similar hybridization structure.
However, it is not always appropriate to expect that nanodiamonds and fulleranes
emit bands at the same wavelength positions.
The fullerane hypothesis  refers to a combination of numerous C$_{60}$H$_m$ isomers with 
different $m$ values, making it sufficiently flexible to match the observational spectrum.

\begin{figure*}
\epsfig{file=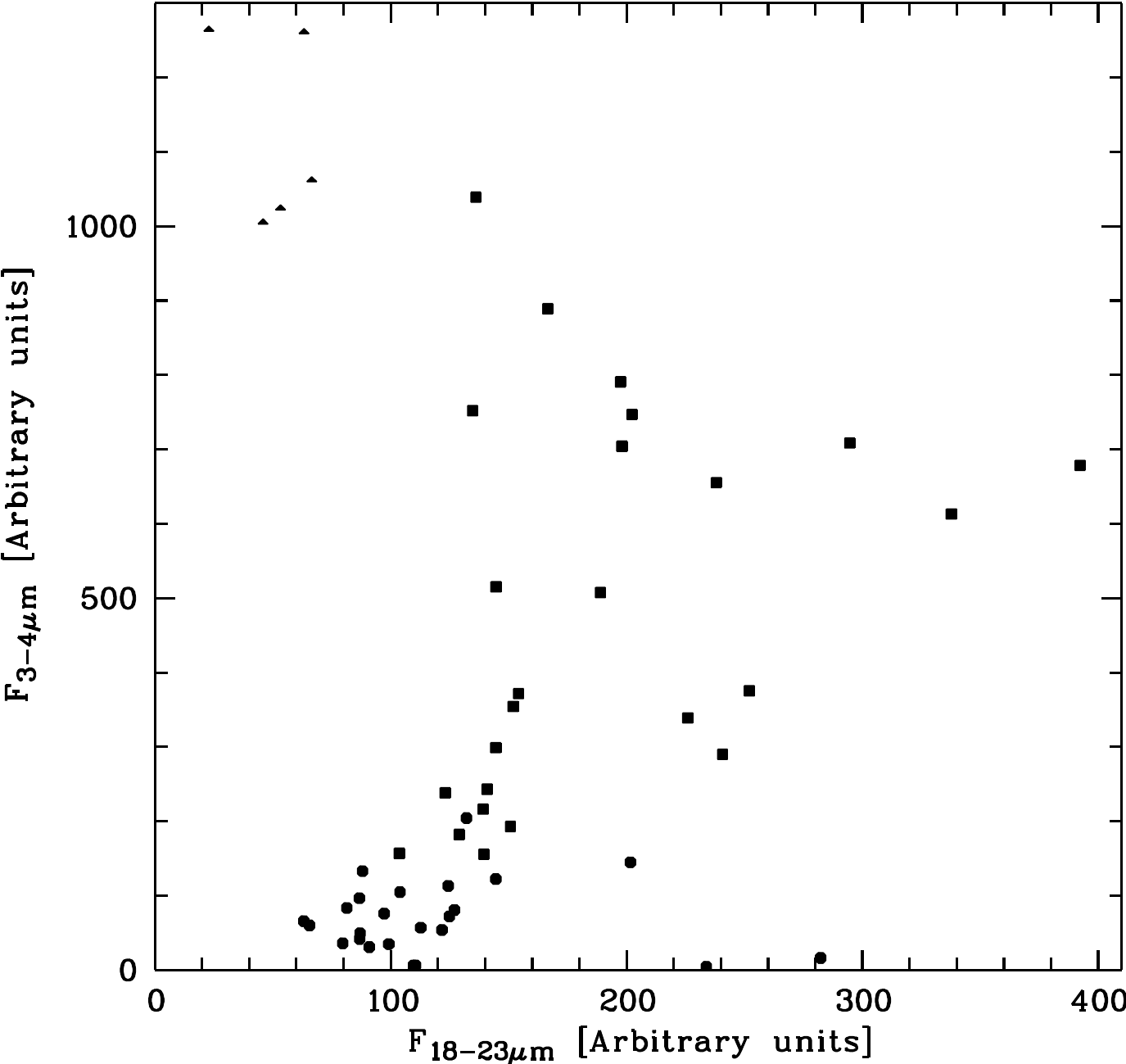, height=10cm}
\caption{Intrinsic strengths of the C$_{60}$H$_{m}$ features lying
in the wavelength range
of 3--4\,$\mu$m versus those of 18--23\,$\mu$m. 
Symbols are the same as those in Figure~\ref{21vsm}.}
\label{3vs21}
\end{figure*}

A significant criteria for  band identification to examine whether associated
subfeatures from the carrier candidate are visible in observed spectra. The C-H stretching mode 
around 3.4\,$\mu$m has been detected in almost all the 21\,$\mu$m sources. However, the theoretical spectra of 
fulleranes imply that there is no correlation between the intensities
of the 3.4\,$\mu$m and the 21\,$\mu$m features, as shown in Figure~\ref{3vs21}. 
The 3.4\,$\mu$m/21\,$\mu$m intensity ratios largely vary among different isomers,
and strong 21\,$\mu$m sources might exhibit weak features at 3--4\,$\mu$. This is conceivable since
the 21\,$\mu$m feature dominantly arise from C-C deformation vibrations. Therefore, the
feature around 3.4\,$\mu$m is not an ideal proxy to test the fullerane hypothesis.

\begin{figure*}
\epsfig{file=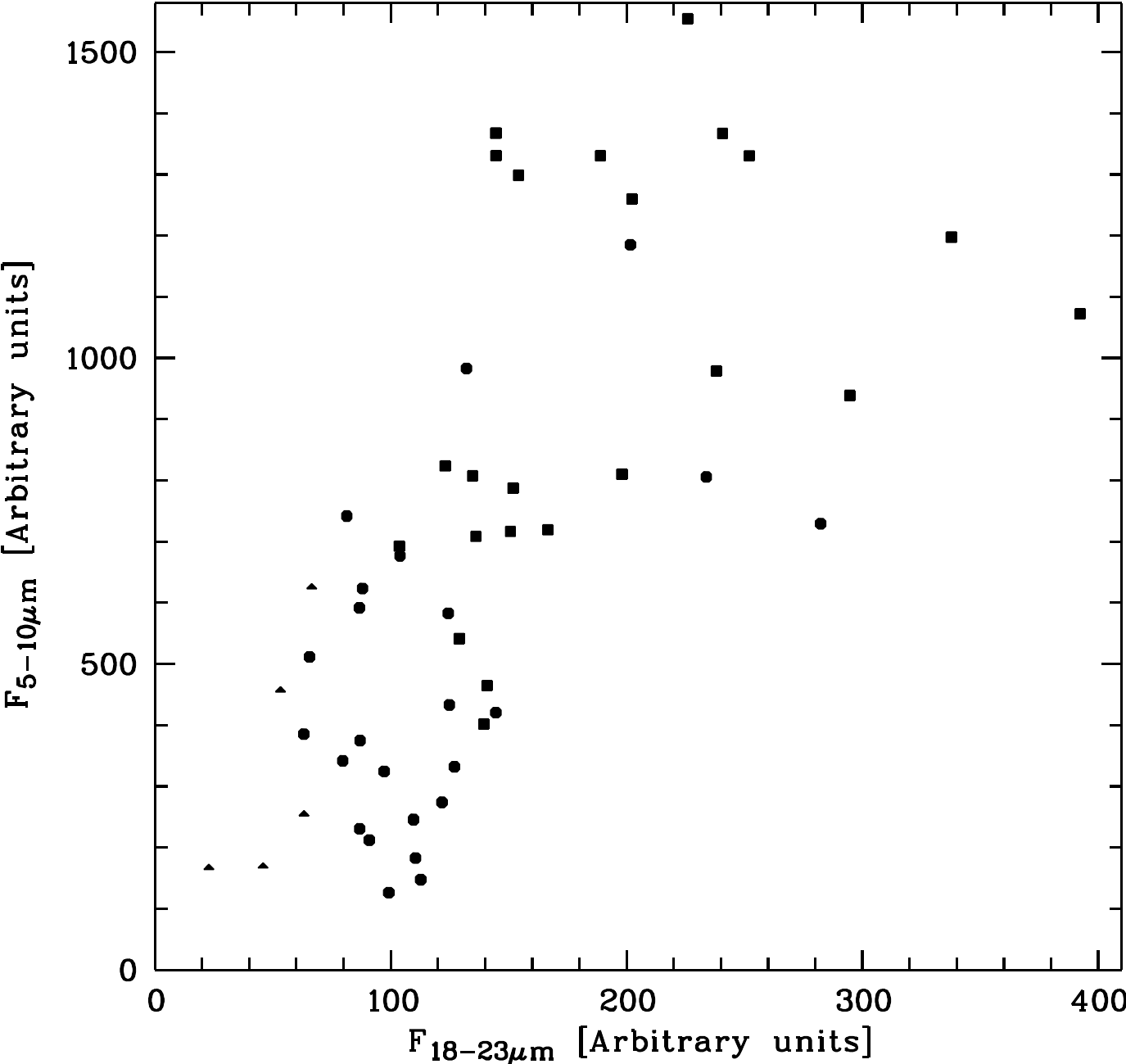, height=10cm}
\caption{Intrinsic strengths of the C$_{60}$H$_{m}$ features lying
in the wavelength range of 5--10\,$\mu$m versus those of 18--23\,$\mu$m. 
Symbols are the same as those in Figure~\ref{21vsm}.}
\label{8vs21}
\end{figure*}

The theoretical spectra of fulleranes reveal that the 5--10\,$\mu$m feature has
an intensity positively correlating with that of the 18--23\,$\mu$m feature
(Figure~\ref{8vs21}).  This makes the fullerane hypothesis plausible since
21\,$\mu$m sources exhibit prominent 5--10\,$\mu$m feature as well.
However, the 5--10\,$\mu$m feature has been commonly detected in various
circumstellar envelopes, and thus it cannot been entirely attributed to fulleranes.

\begin{figure*}
\epsfig{file=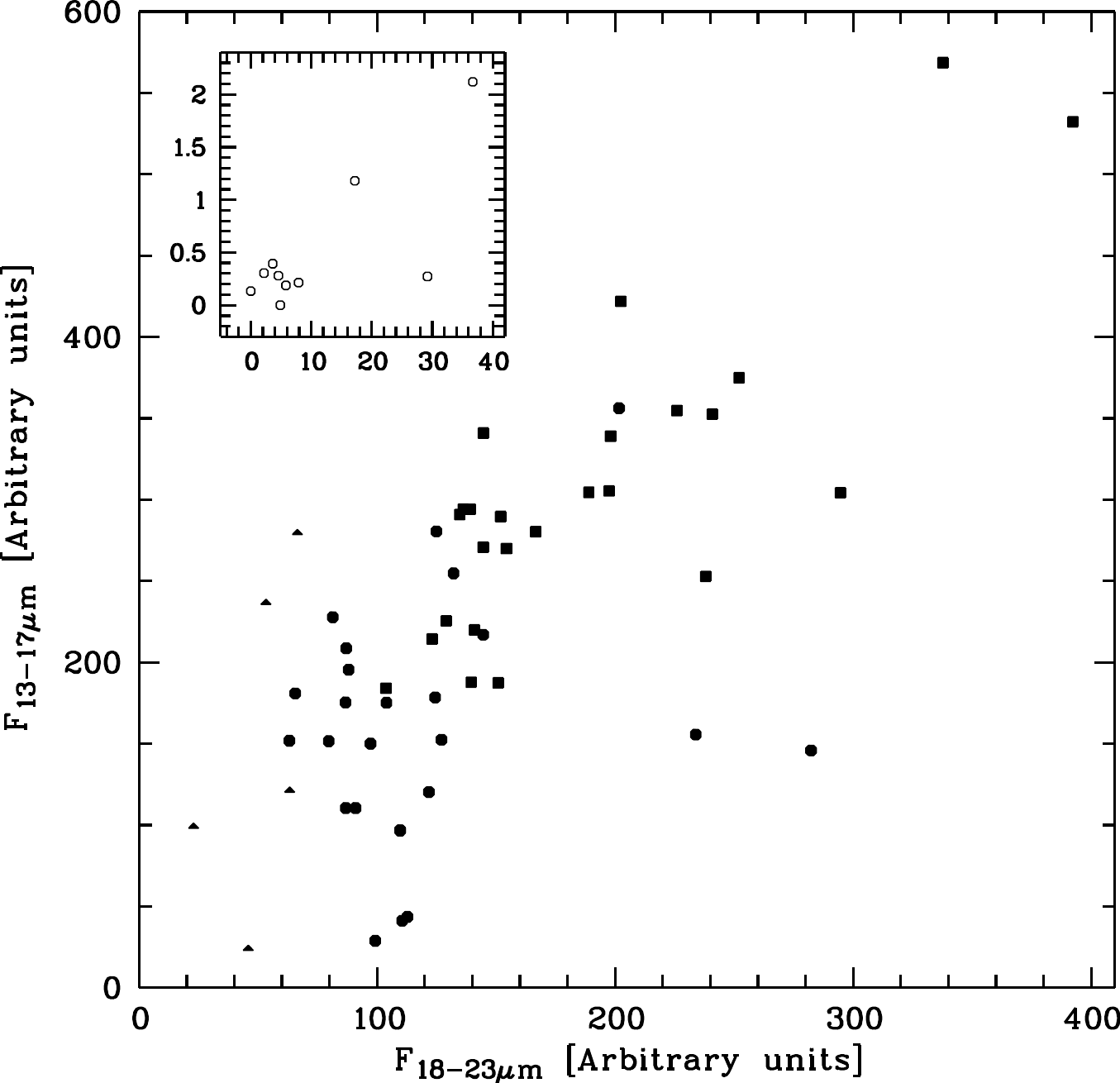, height=10cm}
\caption{Intrinsic strengths of the C$_{60}$H$_{m}$ features lying
in the wavelength range
of 13--17\,$\mu$m versus those of 18--23\,$\mu$m. 
Symbols are the same as those in Figure~\ref{21vsm}.
The insert shows the correlation
between the strengths of the 15.8 $\mu$m the
21 $\mu$m features, as found by \citet{zhang10}. }
\label{15vs21}
\end{figure*}

A critical issue of the fullerane hypothesis is that the 13--17\,$\mu$m feature
arising from fulleranes is hardly visible in the observational spectra.
As shown in Figure~\ref{15vs21}, the 13--17\,$\mu$m feature should have a comparable
intensity with the 21\,$\mu$m one, which is in contrast to the observations.
Nevertheless, it is notable that the  band strengths reproduced by
the DFT calculations are much less accurate than the wavelengths.
The laboratory spectrum of C$_{60}$H$_{18}$ \citep[][see their Figure~3]{ig12}
shows that the 13--17\,$\mu$m feature is much weaker than
the features lying in the 10\,$\mu$m and 18--13\,$\mu$m range.
Moreover, an appropriate excitation model needs to be 
employed to accurately predict the observed intensities.
As a result, the non-detection of the 13--17\,$\mu$m feature 
is insufficient to invalidate the fullerane hypothesis. 
Although no strong band has been detected in the 13--17\,$\mu$m range, a weak feature
at 15.8\,$\mu$m appears in the spectra of all the 21\,$\mu$m sources \citep{hr08}.
\citet{zhang11} found that there is a loose correlation between the intensity of
the 15.8\,$\mu$m and 21\,$\mu$m features (Figure~\ref{15vs21}), suggesting that
the two features may arise from the same material. This gives a seemingly plausible
support for the fullerane hypothesis.

The theoretical spectra of fulleranes do not exhibit strong bands around 30\,$\mu$m.
Therefore, the 30\,$\mu$m feature detected in AGB stars, PPNe, and PNe is unlikely
due to fulleranes. While the 21\,$\mu$m feature, without exception, is accompanied by the
30\,$\mu$m feature, 
the number of the 30\,$\mu$m sources  is much larger, and the strengths of the
two features are not correlated with each other. A common property of the two
features is that both are detected in C-rich environments. Strikingly, 
the 30\,$\mu$m feature was also viewed in the spectra of the PPNe and
PNe detected in C$_{60}$ \citep{zhang13}.  It is tempting to conjecture that  
the carrier of the 30\,$\mu$m feature can be synthesized during the AGB stage,
and then when exposed on UV radiation at PN and PPN stages it is fragmented and partly converted into
fullerenes and fulleranes through a top-down scenario.

\section{Conclusions}

We investigated the theoretical spectra of C$_{60}$H$_{m}$ with the aim of examining 
whether fulleranes are responsible for the 21\,$\mu$m feature in PPNe. 
Based on a comparison of wavelengths and intensities, we infer that moderately 
hydrogenated C$_{60}$ are a promising carrier material producing this feature.
A mixture of specific C$_{60}$H$_{m}$ isomers is required to fit the observed 
profile of the 21\,$\mu$m feature. The survive of fullerane is very sensitive 
to the UV radiative field, and thus can give a natural explanation for
the emergence of the 21\,$\mu$m feature on very short timescales. The main
issue is that the theoretical spectra of fullerane pose strong features in the 13--17\,$\mu$m 
range which are absent in 21\,$\mu$m sources. However, 
given the limits of computaion and model complexity,  this is no enough
to completely rule out the fullerane hypothesis. The analysis and results
proposed here can provide a guideline for future experimental efforts on
the identification of the 21\,$\mu$m feature.


\acknowledgments

This work was supported by National Science Foundation (NSFC). I thank Dr. SeyedAbdolreza Sadjadi
for his help on the computations of C$_{60}$H$_{m}$ spectra.

\clearpage

\end{document}